Fabrication of high performance $MgB_2$ wires by an internal Mg diffusion process

J. M. Hur[1)2)], K. Togano[1)], A. Matsumoto[1)], H. Kumakura[1)]
H. Wada[2)], K. Kimura[2)]

[1)]Superconducting Materials Center, National Institute for materials Science, 1-2-1, Sengen, Tsukuba, Ibaraki 305-0047, Japan
[2)]Department of Advanced Materials Science, The University of Tokyo, Kiban-toh 502, 5-1-5 Kashiwanoha, Kashiwa-shi, Chiba 277-8561, Japan

E-mail: KUMAKURA.Hiroaki@nims.go.jp

**Abstract**

We succeeded in the fabrication of high-$J_c$ $MgB_2$/Fe wires applying the internal Mg diffusion (IMD) process with pure Mg core and SiC addition. A pure Mg rod with 2 mm diameter was placed at the center of a Fe tube, and the space between Mg and Fe tube was filled with B powder or the powder mixture of B-(5mol%)SiC. The composite was cold worked into 1.2mm diameter wire and finally heat treated at temperatures above the melting point of Mg(~650$^o$C). During the heat treatment liquid Mg infiltrated into B layer and reacted with B to form $MgB_2$. X-ray diffraction analysis indicated that the major phase in the reacted layer is $MgB_2$. SEM analysis shows that the density of $MgB_2$ layer is higher than that of usual powder-in-tube(PIT) processed wires. The wires with 5mol% SiC addition heat treated at 670$^o$C showed $J_c$ values higher than $10^5$A/cm$^2$ in 8T and 41,000A/cm$^2$ in 10T at 4.2K. These values are much higher than those of usual PIT processed wires even compared to the ones with SiC addition. Higher density of $MgB_2$ layer obtained by the diffusion reaction is the major cause of this excellent $J_c$ values.

**Introduction**

Because the superconducting transition temperature, $T_c$, of $MgB_2$ is much higher than that of conventional metallic superconductors, $MgB_2$ is expected to be a promising candidate for practical applications. The lower material costs of Mg and B than of Nb are additional advantages of $MgB_2$. $T_c$ values of ~40 K suggest that $MgB_2$ can be used with a convenient cryocooler as a conductor for a cryogen-free magnet at elevated temperatures of ~20 K. For this reason, many studies have already been reported on the fabrication and properties of $MgB_2$ wires.

Currently the powder-in-tube (PIT) process is now most widely used to fabricate $MgB_2$ wires[1-5] and the addition of impurities such as nano-SiC particles has been found to be effective in improving the critical current properties[6]. However, the PIT process is based on powder sintering, therefore, it is difficult to achieve high density $MgB_2$. Thus, the critical current density $J_c$ of PIT processed wires still remains much lower compared to the practical level. In particular in the case of in situ method, in which the powder mixture of Mg and B powders is reacted in a metal sheath, a contraction in volume occurs when the Mg and B react to form $MgB_2$ compound that is responsible for the void formation in the $MgB_2$ layer.

A diffusion process, on the other hand, is effective in achieving higher density $MgB_2$. Giunchi et al reported a successful result of $MgB_2$ wire production using a composite billet composed of a steel pipe internally lined with Nb tube filled with a coaxial internal pure Mg rod and B powder[7]. However, their $J_c$ values are not so high at 4.2K in magnetic fields probably due to the high temperature heat treatment of 850-900°C which brings the decrease of upper critical field[8]. Very recently, they also tried C-doping to the wire and obtained the increase of $B_{irr}$ from 10 to 16T at 4.2K[9]. Togano et al. fabricated $MgB_2$/Fe wires with similar method using Mg-Li bcc alloy tube or rod which is more easily mechanically cold worked than the pure Mg[10-11] and showed that the reacted $MgB_2$ layer has dense structure. However, Li was diffused into $MgB_2$ layer, which deteriorated $J_c$ values of $MgB_2$ layer. This paper reports a significant effectiveness of nano-SiC addition and low temperature heat treatment above the melting point of Mg in order to improve the in-field $J_c$ of the internal Mg diffusion(IMD) processed $MgB_2$ wire using pure Mg. We succeeded in obtaining much higher $J_c$ at 4.2K than those of PIT processed wires.

**Fabrication of wires**

Figure 1 shows schematic illustration of the IMD process. The process is basically the same as the method that employed Mg-Li rod[10]. The sheath material was a pure Fe tube of 6mm outer diameter, 4mm inner diameter and 50mm in length. A pure Mg rod with 2 mm diameter was placed at the center of this Fe tube, and the space between Mg and Fe tube was filled with the amorphous B powder(99.99%, -300 Mesh) or the powder mixture of B and 5 mol%SiC(a few tenths nanometer size). The weight ratio of Mg rod to packed powders was about 3:1 (for B+5mol%SiC mixed powder), which is rather Mg excess compared to the stoichiometric composition. The composite was initially groove-rolled into a rod shape with 2.3 x 2.3 mm$^2$ cross section and then drawn into a wire of 1.2 mm in diameter. No intermediate annealing was required through the cold-working processes.

Figure 2(a) shows optical micrographs of the polished transverse and longitudinal cross sections of the as-drawn 5mol%SiC added wire. The cross sections show uniform deformation of the composite. This is surprising because it is commonly recognized that pure Mg possesses extremely poor deformability at room temperature due to its hexagonal structure limiting the slip plane only in basal plane. The mechanism of improved deformability of pure Mg in the composite is not clear. It is supposed that the configuration in which Mg rod dressed by fine B powders is surrounded by ductile Fe has a beneficial influence on the deformation of Mg rod such as causing cross slip and annihilation of dislocations and recrystallization[12]. After the cold drawing the wire was cut into short pieces of 40 mm in length. The short pieces were wrapped in Zr foils and were heat treated under Ar gas atmosphere. In the case of PIT process, high $J_c$ values at 4.2K are obtained by the low temperature heat treatment at 600~700°C[13]. In IMD process, the liquid Mg infiltration into B layer is necessary before the reaction, therefore heat treatment temperature should be higher than the melting point of Mg, around 650°C. In this study, we heat treated our wires at 670 °C for 3hr, 700 °C for 1h and 800 °C for 1h.

**Characterization of wires**

1. Microstructure

Figure 2(b) shows the optical micrograph of transverse and longitudinal cross sections of the 5 mol%SiC added wire heat treated at 700°C for 1hr. Reacted layer is observed between Mg rod and Fe sheath. During the heat treatment Mg in the liquid state infiltrate into the B layer and react with B to form a layer along the inner wall of the Fe sheath, which is seen as a ring on the cross section. The layer formed by the reaction is almost uniform along the longitudinal direction as shown in the figure. By the consumption of Mg, a large hole is formed at the Mg core which was clearly observed before the heat treatment in Fig.2(a). In some places, Mg layers were observed inside the MgB$_2$ layer, which remained due to the Mg excess nominal composition. The wires heat treated at different temperatures and no SiC added wires also show similar cross sections.

In order to identify the phases formed by the reaction, X-ray diffraction(XRD) analysis was performed for the reacted layer obtained by scraping it off from the Fe sheath after the heat treatment. Figure 3 shows X-ray diffraction pattern of the reacted layers obtained from the 5mol%SiC added wire heat treated at 700°C. Relatively large peaks of Fe are those of Fe sheath material, which was included during scraping off the reacted layer. Except for those Fe peaks, almost all the peaks are identified as $MgB_2$ phase and Mg phase, indicating that major phase formed by the reaction is $MgB_2$. However, some small peaks of impurities such as MgO and $Mg_2Si$ were observed.

Figure 4(a) shows the SEM image of the fractured cross section of $MgB_2$ layer for 5mol%SiC added wire heat treated at 700°C for 1hr. For comparison the $MgB_2$ core of conventional PIT processed wire is also shown in the Fig. 4(b)[11]. The PIT processed wire shows granular $MgB_2$ microstructure, which is typical for PIT processed $MgB_2$. In the IMD-processed wire, on the other hand, the $MgB_2$ layer is more closely packed, and the porosity in $MgB_2$ layer is significantly smaller than that of the PIT processed wire.

2. Superconducting properties

Superconducting transition of the wire was observed by the DC magnetization method using SQUID magnetometer(Quantum Design MPMS). The onset of the transition of the wire without SiC addition heat treated at 700°C for 1h was ~38K which was comparable to that of PIT processed samples heat treated at the same temperature. The 5mol%SiC addition decreased $T_c$ of $MgB_2$ down to ~36K. This result is similar to those of the PIT processed wires[8].

Transport critical current $I_c$ of the wire was measured by the four probe resistive method at 4.2K in magnetic fields up to 12T. Field was applied perpendicularly to the wire axis. Voltage and current leads were directly attached to the Fe sheath surface by soldering. $I_c$ was defined by a potential drop of 1μV appearing across the voltage probes of 10mm distance. The $J_c$($MgB_2$ layer) was calculated for the cross-sectional area of $MgB_2$ layer.

Figure 5 shows transport $J_c$ vs. field curves at 4.2 K of the pure and 5mol%SiC added IMD processed wires. The data of pure and 5 mol% SiC doped wires fabricated by the PIT method are also shown for comparison. Only the data in fields above 6T are shown in the figure because, in the low magnetic fields, lower than 6T, it is difficult to measure the $I_c$ due to the thermal heating of the Fe sheath materials. The IMD processed wire shows higher $J_c$ than the PIT processed tape. The $J_c$ of the IMD processed wire increases with decreasing the heat treatment temperature. Similar result was obtained for PIT processed wire[13]. The 5mol%SiC added IMD wire shows higher $J_c$ and smaller field dependence of $J_c$ than the no SiC added wires, suggesting that some C atoms substitute for B atoms in $MgB_2$ lattice enhancing upper critical field $B_{c2}$ as in the case of PIT processed wire[14]. This is confirmed by the a-axis lattice parameter of 0.3792nm determined by XRD, which is smaller than that of non-doped $MgB_2$. The 5mol%SiC added IMD processed wire heat treated at 670°C shows excellent $J_c$ values. The wire shows $J_c$ above $10^5$ A/cm$^2$ in 8T. At 10T $J_c$ reached 41kA/cm$^2$. We believe that these transport $J_c$ values are the highest ones reported so far for the $MgB_2$ wires. These values are about three times higher than those of the PIT processed wire. The higher $J_c$ values of the IMD processed wires are presumably due to the higher $MgB_2$ layer density as shown in Fig. 4.

**Conclusion**

We succeeded in the fabrication of high-$J_c$ $MgB_2$/Fe wires applying the internal Mg diffusion (IMD) process with pure Mg rod and nano-SiC added B powders. The composite of pure Mg/[B(-SiC) powder]/Fe tube with outer diameter of 6mm was successfully cold worked into 1.2mm wire at room temperature without any breakage. SEM analysis of the heat treated wire clearly indicates that the density of $MgB_2$ layer in the wire was higher than that of a PIT processed wire. $J_c$ of the wire increased with decreasing the heat treatment temperature. $J_c$ of the wire heat treated at 670°C exceeded $10^5$A/cm$^2$ in 8T and 40,000A/cm$^2$ in 10T at 4.2K. These $J_c$ values are much higher than those of usual PIT processed wires. These high $J_c$ values can be attributed to the high density $MgB_2$ layer obtained by this diffusion method. Thus, the densification of $MgB_2$ layer is effective in enhancing $J_c$ of $MgB_2$ wires. Higher $J_c$ values can be expected by the optimization of several parameters such as heat treatment temperature, the ratio of diameter between Mg core and the B powder of starting composite, the cross sectional area reduction of the starting composite, the amount of SiC addition and so on.

References

1) Jin S, Mavoori H and Dover RB 2001 *Nature* **411** 563-565.
2) Kumakura H, Matsumoto A, Fujii H and Togano K 2001 *Appl. Phys. Lett.* **79** 2435-37.
3) Flukiger R, Lezza P, Beneduce C, Musolino N and Suo HL 2003 *Supercond. Sci. Technol.* **16** 264-270.
4) Fujii H, Togano K and Kumakura H 2002 *Supercond. Sci. Technol.* **15**, 1571-76.
5) Sumption MD, Wu X, Rindfleisch M, Bhatia M, Wu X, Tomsic M and Collings EW 2005 *cond. Sci. Technol.* **18** 730-734.
6) Dou X, Soltanian S, Horvat J, Wang XL, Zhou SH, Ionescu M, Liu HK, Munroe P and Tomsic M 2002 *Appl. Phys. Lett.* **81** 3419.
7) Giunchi G, Ceresara S, Ripamonti G, Zenobio A Di, Rossi S, Chiarelli S, Spadoni M, Wesche R and Bruzzone PL 2003 *Supercond. Sci Technol.* **16** 285-291
8) Kumakura H, Kitaguchi H, Matsumoto A and Yamada H 2005 *Supercond. Sci. Technol.* **18** 1042-46
9) Giunchi G, Lipamonti G, Perini E, Cavallin T, and Bassani E 2007 *IEEE Trans. Appl. Supercond.* **17** 2761-65
10) Togano K, Fujii H, Takeuchi T and Kumakura H 2007 *Supercond. Sci. Technol.* **20** 629-633.
11) Togano K, Fujii H, Takeuchi T, Kumakura H 2007 *Supercond. Sci. Technol.* **20** 239-243.
12) Jotoku M, Yamamoto A and H. Tsubakino H 2006 *Journal of Japan Institute of Light Metals*(in Japanese) **56** 711-715.
13) Fujii H, Togano K and Kumakura H 2002 *Supercond. Sci. Technol.* **15** 1571-76
14) Yamada H, Hirakawa M, Kumakura H and Kitaguchi H 2006 *Supercond. Sci. Technol.* **19** 175-177

Figure Captions

Figure 1. Schematic illustration of the $MgB_2$/Fe composite wire fabrication process by the internal Mg diffusion method.

Figure 2. Optical micrographs of the transverse and longitudinal cross sections of the 5mol%SiC added wires. a) As-drawn wire. b) After the heat treatment at 700$^o$C for 1hr.

Figure 3. X-ray diffraction pattern of the reacted layers obtained from the 5mol%SiC added wire heat treated at 700$^o$C.

Figure 4. SEM images of the fractured cross section of $MgB_2$ layer in the wires. a)5mol%SiC added IMD processed wire heat treated at 700$^o$C for 1hr. b) PIT processed wire.

Figure 5. Transport $J_c$ vs. field curves at 4.2 K of the pure and 5mol%SiC added IMD processed wires. The data of PIT processed wires are also shown for comparison.

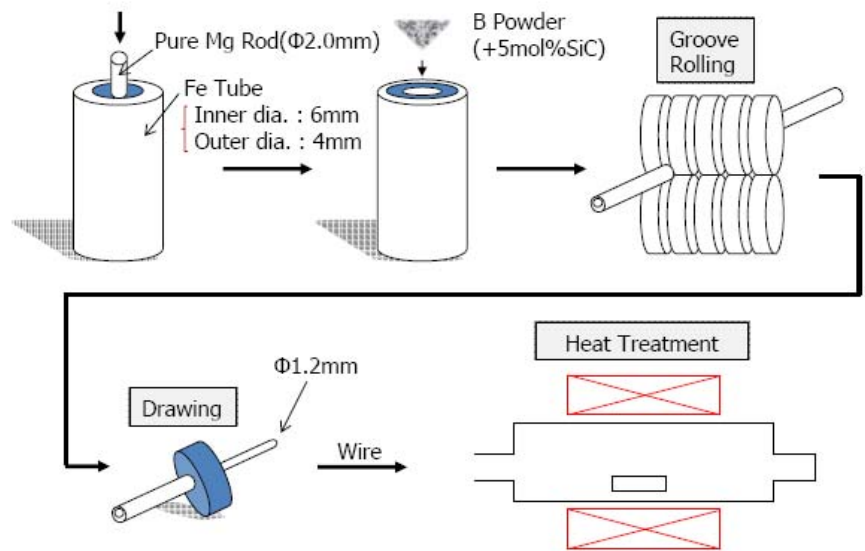

Figure 1. Schematic illustration of the MgB$_2$/Fe composite wire fabrication process by the internal Mg diffusion method.

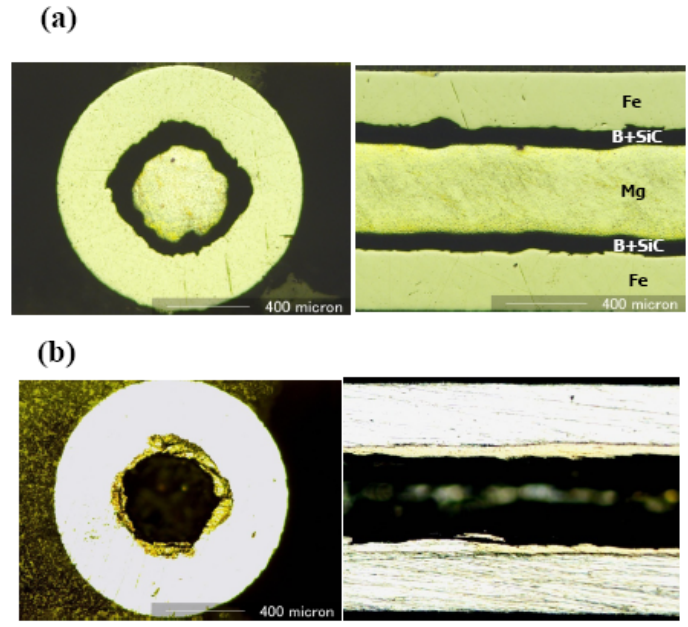

Figure 2. Optical micrographs of the transverse and longitudinal cross sections of the 5mol%SiC added wires.  a) As-drawn wire.  b) After the heat treatment at 700°C for 1hr.

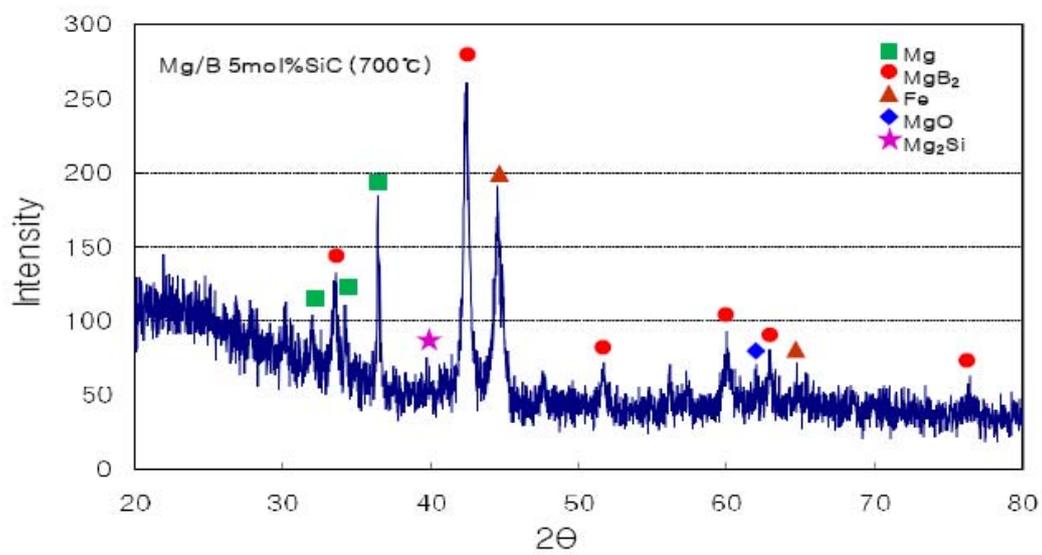

Figure 3. X-ray diffraction pattern of the reacted layers obtained from the 5mol%SiC added wire heat treated at 700°C.

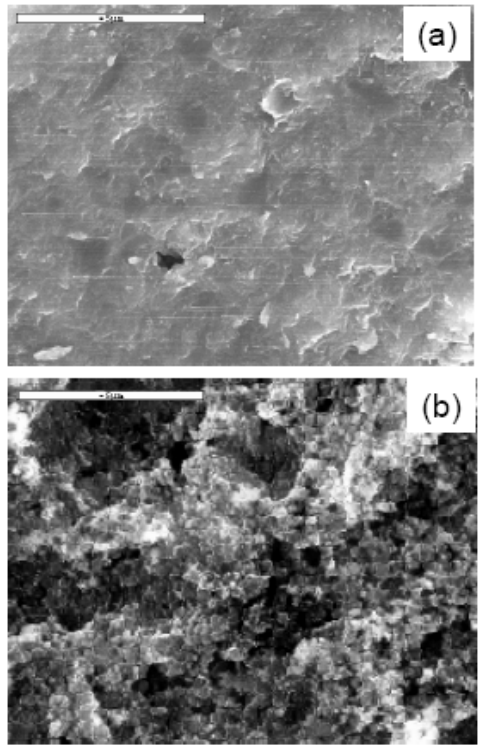

Figure 4. SEM images of the fractured cross section of $MgB_2$ layer in the wires.
a) 5mol%SiC added IMD processed wire heat treated at 700°C for 1hr.
b) PIT processed wire.

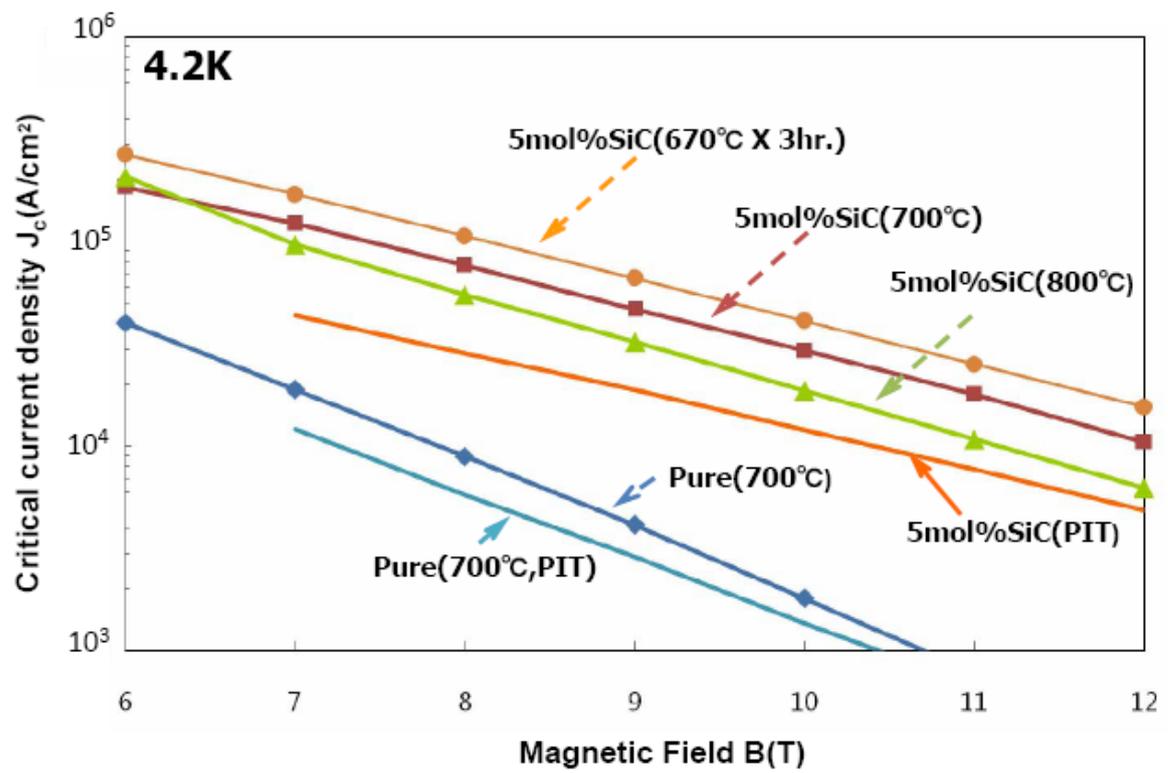

Figure 5. Transport $J_c$ vs. field curves at 4.2 K of the pure and 5mol%SiC added IMD processed wires. The data of PIT processed wires are also shown for comparison.